\documentclass[10pt,twocolumn,letterpaper]{article}

\usepackage{cvpr}
\usepackage{times}
\usepackage{epsfig}
\usepackage{graphicx}
\usepackage{amsmath}
\usepackage{amssymb}
\usepackage{float}
\usepackage{subfigure}
\usepackage{multirow}
\usepackage{authblk}

\usepackage[breaklinks=true,bookmarks=false]{hyperref}

\cvprfinalcopy 

\def\cvprPaperID{****} 

\begin{document}

\def\andname{,}
\author[1]{Jing Zhou\thanks{Corresponding author}}
\author[1]{Sihan Wen}
\author[2]{Akira Nakagawa}
\author[2]{Kimihiko Kazui}
\author[1]{Zhiming Tan}
{
	\makeatletter
	\renewcommand\AB@affilsepx{, \protect\Affilfont}
	\makeatother	
	\affil[1]{Fujitsu R\&D Center Co. Ltd. Shanghai, China}
	\affil[2]{Fujitsu Laboratories Ltd. Kawasaki, Japan}
}
{
	\makeatletter
	\let\AB@affilsep\AB@affilsepx
	\makeatother	
	\affil[1]{\small \{zhoujing, wensihan, zhmtan\}@cn.fujitsu.com,}
}
\affil[2]{\small \{anaka, kazui.kimihiko\}@fujitsu.com}
\title{Multi-scale and Context-adaptive Entropy Model for Image Compression}
\renewcommand\Authands{, }

\maketitle

\begin{abstract}
   We propose an end-to-end trainable image compression framework with a multi-scale and context-adaptive entropy model, especially for low bitrate compression. Due to the success of autoregressive priors in probabilistic generative model, the complementary combination of autoregressive and hierarchical priors can estimate the distribution of each latent representation accurately. Based on this combination, we firstly propose a multi-scale masked convolutional network as our autoregressive model. Secondly, for the significant computational penalty of generative model, we focus on decoded representations covered by receptive field, and skip full zero latents in arithmetic codec. At last, according to the low-rate compression's constraint in CLIC-2019, we use a method to maximize MS-SSIM by allocating bitrate for each image.
\end{abstract}

\section{Introduction}

Recently, artificial neural networks have emerged as a promising direction and achieved many breakthroughs. Image compression is a fundamental and well-studied technique in past decades. The key challenge is to control trade-off between two competing costs: entropy of discretized representation (rate) and error arising from quantization (distortion). Models related to autoencoder \cite{balle2016end,balle2018variational,theis2017lossy,lee2018context}, RNN \cite{RNN}, and GAN \cite{agustsson2018generative,rippel2017real} were proposed to achieve joint optimization of rate and distortion. These methods have got great success, and some of them have surpassed successful codecs such as JPEG, JPEG2000, and BPG. 

In the rate-distortion optimization $R + \lambda\cdot D$, where $\lambda$ acts as a balance between the rate ($R$) and the distortion ($D$). For the distortion, MSE (Mean Square Error) / PSNR is widely used. Nowadays it can also be measured with Multi-Scale Structural SIMilarity (MS-SSIM), especially in deep learning methods. According to information theory, the rate can be estimated by an entropy model. Because the actual distributions of latent representations are unknown, the entropy model should learn to estimate probabilistic distribution. So the most important part is a trainable and accurate entropy model, which can represent the rate explicitly. To predict probability model for each representation, Ball{\'e} et al. \cite{balle2018variational} , Theis et al. \cite{theis2017lossy}, Mentzer et al. \cite{mentzer2018conditional} proposed novel and input-adaptive frameworks for entropy model. 

Our proposed framework is based on Minnen et al. \cite{minnen2018joint} to exploit an accurate probabilistic structure for latents. We mainly focus on an entropy model with complementary combination of autoregressive and hierarchical priors. Each representation is modeled with a Gaussian distribution, and all parameters of the distribution are predicted one by one. Then two methods are presented by considering the trade-off between performance and speed. The first one is to reduce redundant computation. The second is to ignore full zero feature maps in latents while using arithmetic codec. At last, considering the bitrate constraint, a method of bit allocation for each image is employed to pursue better performance on MS-SSIM.

\section{The proposed framework}

\begin{figure*}[t]
	\begin{center}
		\includegraphics[width=0.7\linewidth]{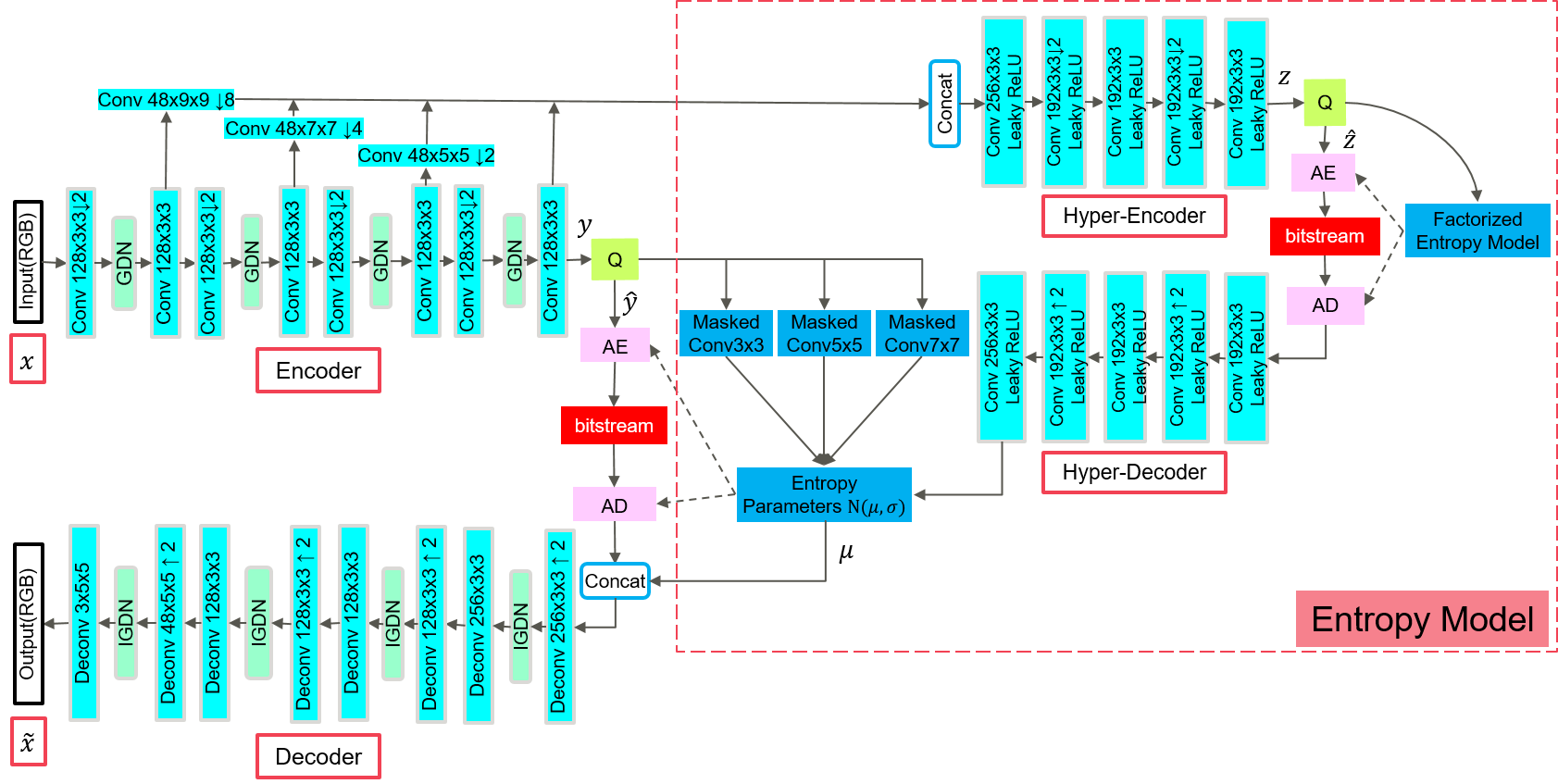}
	\end{center}
	\caption{Our framework. Conv: Convolution layer. $128\times 3\times3$, 128: number of feature map, $3\times3$: kernel height $\times$ width. $\downarrow 2$: downsampling with stride 2; $\uparrow 2$: upsampling with stride 2. Masked Conv $3\times 3$: Masked convolution with $3\times 3$ kernel \cite{van2016conditional}. Deconv: Deconvolutional layer. GDN: Generalized Divisive Normalization; IGDN: Inverse GDN \cite{balle2016end}. Q: Quantization; AE: Arithmetic Encoder; AD: Arithmetic Decoder.}
	\label{fig:1}
\end{figure*}

The whole framework is shown in Figure \ref{fig:1}, which can be briefly divided into two parts. The first one is an asymmetric autoencoder network. It transforms original image \emph{x} from pixel-level to high-level representations with \emph{Encoder} and reconstructs them back to \emph{$\tilde{x}$} with \emph{Decoder}. The second is an \emph{Entropy Model}, which mainly contains a hyper autoencoder and an autoregressive model with three masked convolutional layers. The \emph{Entropy Parameters} is made up of several $1\times1$ convolutional layers as Minnen et al. \cite{minnen2018joint}.  A \emph{Factorized Entropy Model} is used for $\hat{z}$, which is a fixed and fully factorized prior proposed by Ball{\'e} et al. \cite{balle2018variational}. Assuming a Gaussian distributed probability mass function for $\hat{y}$, the parameters of $\mu$ and $\sigma$ are predicted which are used in arithmetic codec (\emph{AE} and \emph{AD}). Latent representations with real-value are quantized (\emph{Q}) to create $\hat{y}$ and $\hat{z}$ in evaluation, which can be compressed into bitstream. 

\subsection{Entropy model}
In our proposed framework, the side information from hyper prior and context model plays an important role in entropy model. We improve the entropy model's performance from two aspects. 

The first aspect is to extract multi-scale and extended feature maps from intermediate layers in \emph{Encoder}, and bigger inputs are convoluted with larger kernels. Assuming \emph{x}'s shape is $H\times W\times C$, the first extended feature maps are obtained via a $9\times9 \downarrow 8$ convolution  with a $\frac{H}{2}\times \frac{W}{2}\times 128$ input. With an input of $\frac{H}{4}\times \frac{W}{4}\times 128$, the second extended feature maps are obtained via a $7\times7 \downarrow 4$ convolution. The last one can be obtained in the same manner. By fusing these feature maps with latents $\emph{y}$, the input to the \emph{Hyper-Encoder} is obtained. Since more features are added, the number of feature map $z$ is increased to 192, compared with $\hat{y}$ (128). As the outputs of \emph{Hyper Decoder} represents the distribution of $\hat{y}$ roughly, such fusion is beneficial to probability estimation in higher precision.  

\begin{figure}[H]
	\centering
	\subfigure[Scheme of context model]{
		\label{Scheme of context model}
		\includegraphics[width=0.4\linewidth]{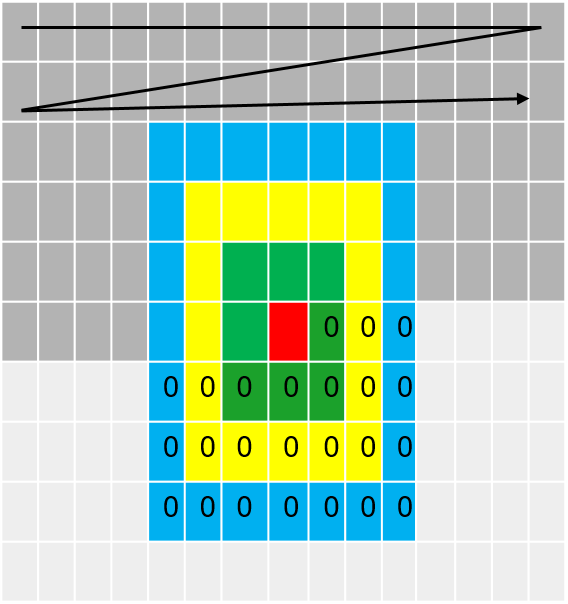}}
	\subfigure[Masks with different kernels]{
		\label{Masks with differernt kernels}
		\includegraphics[width=0.5\linewidth]{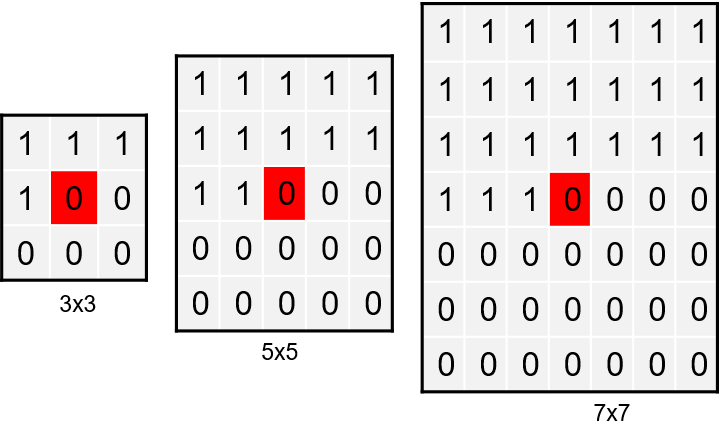}}
	\caption{Multi-scale context model}
	\label{fig:2}
\end{figure}

The second aspect is the autoregressive model. As shown in Figure \ref{fig:2} (a), all points are encoded/decoded in scan order (indicated by the arrow) one by one. To decode the current point (colored with red), only previously decoded points can be used. We propose a multi-scale context model, which contains 3 parallel masked convolutional layers shown in Figure \ref{fig:2} (b). The available information used is the previous decoded points, and the un-decoded ones are masked with zero. Combining with these 3 masked layers, the scope can be divided into 3 rings in Figure \ref{fig:2} (a). The first ring is colored with green, the second colored with yellow and the third blue. With three kernels centered on the current point, all kernels are effective in the first ring; two kernels ($7\times7$ and $5\times5$) in the second ring; only $7\times7$ kernel in the third. With such multi-scale convolutional layers, the influence of points in the closer ring is amplified. 

With the predicted parameters of $\mu$ and $\sigma$, the discrete representation's probability is calculated with Eq \ref{eq-1}, where $\mathcal N(\mu,\sigma^2)$ represents the assumed Gaussian distribution.
\begin{equation}
p(\hat{y}|\hat{z})=\prod_{i}(\mathcal N(\mu_i,\sigma_i^2) \ast \mu_i(-\frac{1}{2}, \frac{1}{2}))(\hat{y}_i)
\label{eq-1}
\end{equation}

So the total bitrate contains two parts: rate $R_{\hat{y}}$ for representation $\hat{y}$ and rate $R_{\hat{z}}$ for side information from hyper-prior $\hat{z}$, as shown in Eq \ref{eq-2}.          
\begin{equation}
R = \underbrace{\sum(-log_2(p(\hat{y}|\hat{z}))}_{R_{\hat{y}}} + \underbrace{\sum-log_2(p(\hat{z}))}_{R_{\hat{z}}}
\label{eq-2}
\end{equation}

\subsection{Adjust quantization error}
The quantization, such as round function, is not applicable in the end-to-end training, because of the problem of zero gradient. To solve it, we adopt the method of noise-based relaxation proposed by Ball{\'e} et al.\cite{balle2016end} in training.

Another problem for the quantization is that it introduces error, which will decrease the performance of reconstruction. As in Eq \ref{eq-1}, the whole framework is trained to minimize the difference between $\hat{y}$ and $\mu$. The predicted $\mu$ in continuous value can supplement some information to the discrete $\hat{y}$. Concatenating $\hat{y}$ and $\mu$ as the input of the \emph{Decoder} can adjust quantization error to some extent. 

\section{Experiments}
\subsection{Training method}
We train our network with more than 6000 images. Our dataset mainly contains three parts: training datasets provided by CLIC, DIV2K, and Flicker2K dataset \cite{Lim_2017_CVPR_Workshops}. We randomly crop patches of 256x256 from the full resolution images for each batch while training. 

In the rate-distortion optimization, the full loss function is shown in Eq \ref{eq-3}. We train our model from scratch in three stages progressively from high bitrate to low bitrate. Firstly, MSE: $\|x-\tilde{x}\|^2$ is used as the distortion (\emph{D}). A stable model is trained with bigger $\lambda$, which performs well in PSNR. Secondly, we switch to MS-SSIM: $D=1 - L_{MS{\text -}SSIM}$. We train the model for better performance in the metric of MS-SSIM. Finally, $w_{1} \cdot (\left|\hat{y}-\mu\right|)$ is added to the loss function, which is beneficial to adjusting the quantization error, where $w_{1}$ is 0.2. 

\begin{equation}
Loss = R + \lambda \cdot D
\label{eq-3}
\end{equation} 

In the MS-SSIM metric, the image is scaled five times by a factor of 2. Then five SSIM values can be obtained. The MS-SSIM is the sum of five weighted SSIMs. We train the models with two different weights: default [0.0448, 0.2856, 0.3001, 0.2363, 0.1333], and average [1.0, 1.0, 1.0, 1.0, 1.0]. After training, we evaluate these two models' performance on validation dataset of CLIC, and the results are shown in Table \ref{tab-1}. We can find that the average weights perform better on PSNR, but worse on MS-SSIM. Because the default weights are both used in the training and evaluation, the default weights' MS-SSIM value is higher. We also think there exists a trade-off between MS-SSIM and PSNR by viewing this result.
\begin{table}[H]
	\footnotesize
	\begin{center}
		\begin{tabular}{|c|c|c|c|}
			\hline
			Weights & MS-SSIM & PSNR & BPP \\
			\hline
			Average & 0.9743 & 30.13 & 0.148 \\
			\hline
			Default & 0.9751 & 29.75 & 0.149 \\
			\hline
		\end{tabular}
	\end{center}
	\caption{Evaluation results on CLIC validation dataset}
	\label{tab-1}
\end{table}
\subsection{Speed up autoregressive model}
Due to the autoregressive network's inherent serial scheme, it's time consuming from practical standpoint especially for big input. Current popular acceleration techniques are in the way of parallelization, which is not suitable in our scheme. To accelerate our model, two methods are proposed.
\begin{table*}[t]
	\footnotesize
	\begin{center}
		\begin{tabular}{|c|c|c|c|c|c|}
			\hline
			Method & Image size & Decoding time(s) & MS-SSIM & PSNR & Entropy(byte)\\
			\hline 
			\multirow{3}*{Before} & 365*512 & 50 & 0.9687 & 34.61 & 2763\\
			\cline{2-6}
			~   & 1448*972 & 2868 & 0.9571 & 27.23 & 53478 \\
			\cline{2-6}
			~   & 2000*2000 & 22275 & 0.9753 & 29.27 & 98865 \\
			\hline 
			
			\multirow{3}*{After}  & 365*512 & 10$(\downarrow 80\%$) & 0.9687 & 34.61 & 2778$(\uparrow 5.4\times 10^{-3}$) \\
			\cline{2-6}
			~     & 1448*972 & 93$(\downarrow 96.76\%$) & 0.9571 & 27.23 & 53493$(\uparrow 2.8\times 10^{-4}$) \\
			\cline{2-6}
			~    & 2000*2000 & 259$(\downarrow 98.84\%$) & 0.9753 & 29.27 & 98879$(\uparrow 1.4\times 10^{-4}$) \\
			\hline 
		\end{tabular}
	\end{center}
	\caption{Performance comparison of our acceleration method}
	\label{tab-2}
\end{table*}

The first method is reducing unnecessary computation in the context model. Assuming $\hat{y}$'s shape is $h\times w\times c$, it means there are $c$ feature maps, and each shape is $h\times w$. The max receptive field for our context model is $7\times 7$. So cropping $7\times 7 \times c$ centered on the point to be decoded is enough. With this operation, the computation is decreased from $h\times w\times c$ to $7\times 7 \times c$ each time. In addition, there is no performance penalty because no information is lost.

The second one is about arithmetic codec. As an intuition, better performance can be obtained if there are more feature maps in the bottleneck. After several trials, 128 feature maps for $\hat{y}$ are proved to be the best choice for low-rate compression. Looking into these 128 feature maps, almost half of them are full-zero. We employ $c$ bits to indicate whether the feature map is full-zero or not and skip points of full-zero feature map while using entropy codec. So extra 128/8 bytes are consumed to store these flags. The pseudocode of selective arithmetic codec is illustrated as below.
\begin{table}[H]
	\footnotesize
	\begin{center}
		\begin{tabular}{l}
			\hline
			Algorithm: selective arithmetic codec \\ \hline
			Input: feature map of $\hat{y}[h,w,c]$ \\ 
			Output: bitstream \\
			1: $flags$ = zeros($c$) \\
			2: for $i$ in range($c$) \\
			3: \quad if sum$(\left|\hat{y}[:,:,i]\right|) > 0 $ \\
			4: \qquad $flags[i]$ = 1  \\
			5: \quad else: \\
			6: \qquad $flags[i]$ = 0 \\
			7: for $h\_idx$ in range($h$) \\
			8:\quad for $w\_idx$ in range($w$) \\
			9:\qquad for $ch\_idx$ in range($c$) \\
			10:\qquad \quad if $flags[ch\_idx)]$ == 1 \\
			11:\qquad \qquad arithmetic-codec($\hat{y}$[$h\_idx$,$w\_idx$,$ch\_idx$]) \\
			\hline			      
			\end {tabular}
		\end{center}
	\end{table}

The decoding is tested under docker environment with 2 CPU cores, and the processor is Intel i7-4790K CPU, 4.00GHz. Three images of different shapes are chosen from CLIC 2019 test dataset. From Table \ref{tab-2}, although entropy increases by a very small margin, our method save time a lot without performance degradation, especially for large images (more than 96\% decoding time).

\subsection{Bit allocation under limited bitrate}
The task of the CLIC is to maximize metrics such as PSNR, MS-SSIM under given bitrate, and total test dataset is seen as a target. It's hard for a model trained with a single $\lambda$ to satisfy this constraint for a random dataset. So multiple models with different bitrates are needed. To maximum performance of the whole test dataset, a knapsack solver is used to allocate each image with appropriate bitstream from these models.

\subsection{Performance}

\begin{figure}[H]	
	\centering
	\includegraphics[width=0.7\linewidth]{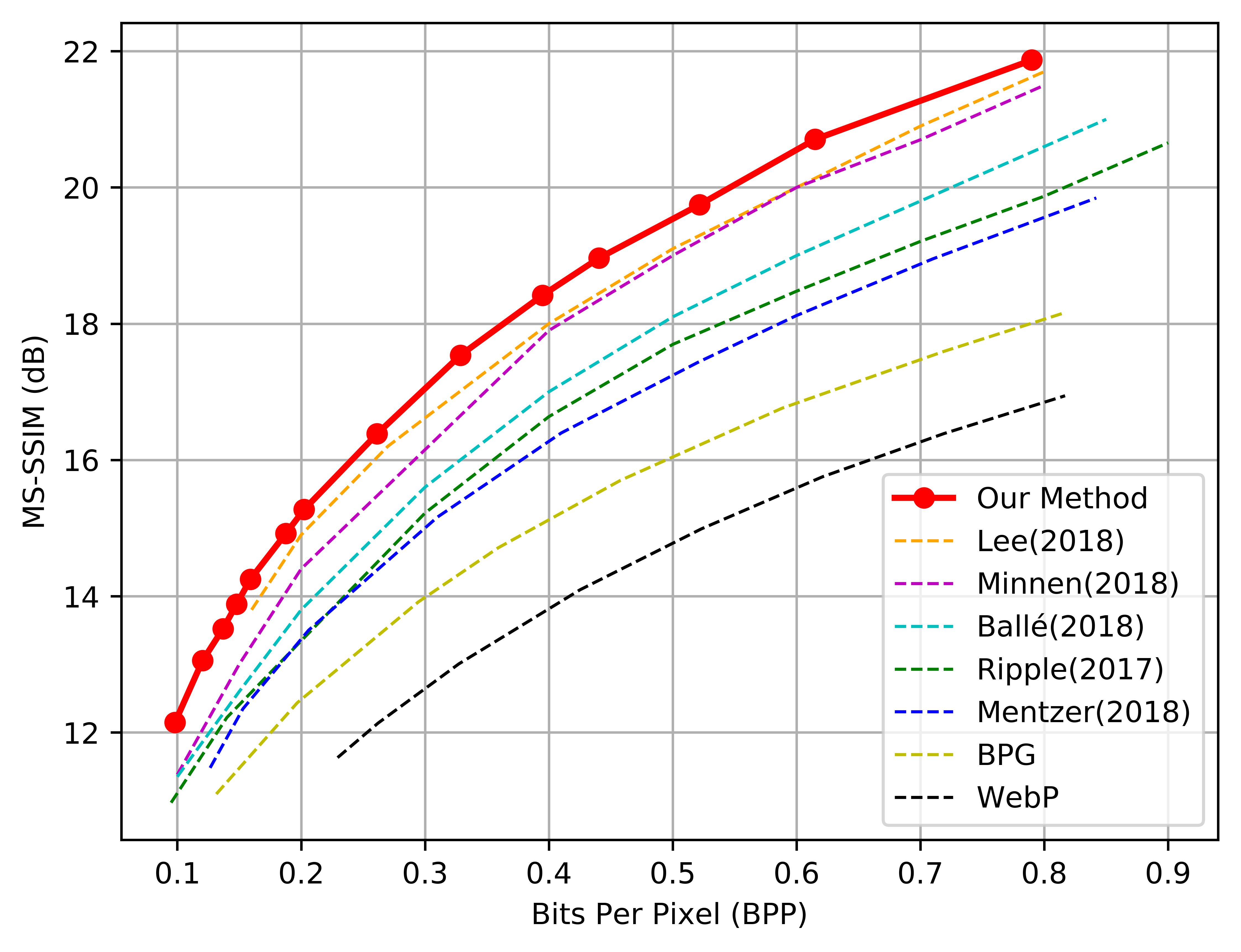}
	\caption{Comparison of Rate-Distortion curves on Kodak}
	\label{fig:3}
	\centering
\end{figure}
We evaluate compression performance on the public Kodak dataset, and rate-distortion curves are shown in Figure \ref{fig:3}. To improve legibility, the MS-SSIM scores are in dB: ${MS{\text -}SSIM}_{dB} = -10\cdot log_{10}(1 - {MS{\text -}SSIM})$. As far as we know, our method proves to be the state-of-the-art compared with other baseline methods.

\begin{table}[H]
	\footnotesize
	\begin{center}
		\begin{tabular}{|c|c|c|c|c|}
			\hline	
			Model number & Rate range & MS-SSIM & PSNR & BPP\\
			\hline
			1 & 0.148 & 0.9721 & 28.51 & 0.148 \\
			\hline			
			6(JointSSIM)& [0.135, 0.163] & 0.9729 & 28.54 & 0.150 \\
			\hline
			8(Joint)& [0.120, 0.180] & 0.9733 & 28.54 & 0.150 \\
			\hline			
			8     & [0.120, 0.267] & 0.9739 & 28.50 & 0.150\\
			\hline
		\end{tabular}
	\end{center}
	\caption{Results on CLIC 2019 test dataset}
	\label{tab-3}
\end{table}

As total bitrate should be no more than 0.15 bpp for CLIC's low-rate task, evaluation results for the test dataset are shown in Table \ref{tab-3}. The \emph{Model number} represents the number of model used with different bitrates. For the \emph{Rate range}, e.g., [0.135, 0.613], 0.163 represents the biggest bpp evaluated on the test dataset and 0.135 represents the lowest. For a single bitrate model, the best performance of MS-SSIM is 0.9721 with 0.148 bpp. With bit allocation method, the value of MS-SSIM can be improved if there is a larger range of bitrate. Our submitted versions are 'JointSSIM' and 'Joint'. For 'Joint' team, it achieves the second place in MS-SSIM and third place in MOS. We further enlarge the range of bit rate, e.g., [0.120, 0.267], and it performs the best in Table \ref{tab-3}.

\section{Conclusion}
In this paper, we propose an end-to-end image compression framework, which can be seen as an extended work of Minnen et al.\cite{minnen2018joint}. We implement our code based on the open source code provided by Johannes Ball{\'e} at \url{https://github.com/tensorflow/compression}. Firstly, we propose a multi-scale and context-adaptive entropy model. Secondly, methods are proposed to accelerate in entropy codec. Lastly, we use a method for bitrate allocation to maximize MS-SSIM. In the future, we will focus more on speeding up our model with improved performance. 

{
	\scriptsize
	\bibliographystyle{ieee_fullname}
	\bibliography{finalbib}
}
\end{document}